\documentclass[twocolumn,showpacs,superscriptaddress,preprintnumbers,floats,aps,pra,10pt]{revtex4-1}
\usepackage{graphicx}
\usepackage{subfigure}
\usepackage{dcolumn}
\usepackage{amsmath}
\usepackage{amsfonts}
\usepackage{amssymb}
\usepackage{color}
\usepackage{ulem}
\providecommand{\U}[1]{\protect\rule{.1in}{.1in}}

\begin{document}
\title{Testing dissipative dark matter in causal thermodynamics}
\author{Norman Cruz}
\altaffiliation{norman.cruz@usach.cl}
\affiliation{Departamento de F\'isica, Universidad de Santiago de Chile, \\
Avenida Ecuador 3493, Santiago, Chile.}
\author{Esteban Gonz\'alez}
\altaffiliation{esteban.gonzalezb@usach.cl}
\affiliation{Departamento de F\'isica, Universidad de Santiago de Chile, \\
Avenida Ecuador 3493, Santiago, Chile.}
\author{Guillermo Palma}
\altaffiliation{guillermo.palma@usach.cl}
\affiliation{Departamento de F\'isica, Universidad de Santiago de Chile, \\
Avenida Ecuador 3493, Santiago, Chile.}
\date{\today}

\begin{abstract}

 In this paper we study the consistency of a cosmological model representing a universe filled with a one-component dissipative dark matter fluid, in the framework of the causal Israel-Stewart theory, where a general expression arising from perturbation analysis for the relaxation time $\tau$ is used. This model is described by an exact analytic solution recently found in [N. Cruz, E. Gonz\'alez and G. Palma, Gen. Rel. Grav. \textbf{52}, 62 (2020), which depends on several model parameters as well as integration constants, allowing the use of Type Ia Supernovae and Observational Hubble data to perform an astringent observational test. The constraint regions found for the parameters of the solution allow the existence of an accelerated expansion of the universe at late times, after the domination era of the viscous pressure, which holds without the need of including a cosmological constant. Nevertheless, the fitted parameter values lead to drawbacks as a very large non-adiabatic contribution to the speed of sound, and some inconsistencies, not totally conclusive, with the description of the dissipative dark matter as a fluid, which is nevertheless a common feature of these kind of models.

\end{abstract}

\vspace{0.5cm}
\maketitle

\section{Introduction}
It is well accepted that nowadays the cosmological data consistently
indicates that the expansion of the Universe began to
accelerate~\cite{Riess, Perlmutter, WMAP, Planck, Planck2016} around
$z=0.64$~\cite{Moresco}. Thus, every model used to describe the
cosmic background evolution must display this transition in its
dynamics. Of course, $\Lambda$CDM presents this transition as well
and it can be understood as the transition between the dark matter
(DM) dominant era and the era dominated by the dark energy (DE).
Nevertheless, despite the fact that the $\Lambda$CDM model has been
very successful to explain the cosmological data, it presents the
following weak points from the theoretical point of view: i) Why the
estimated value of $\Lambda$  is 120 orders of magnitude smaller than
the physically predicted one?. This is the well known cosmological constant
problem~\cite{Weinberg, Carroll, Turner, Sahni, Carroll2001,
Padmanabhan2003, Peebles}, which can be represented mainly by the
two following open questions: a) Why does the observed vacuum energy
has such an unnaturally small but non vanishing value?, and b) Why
do we observe vacuum density to be so close to matter density, even
though their ratio can vary up to 120 orders of magnitude during the
cosmic evolution? (known as the coincidence problem)~\cite{Steinhardt,
Zlatev}. This model presents serious specific observational
challenges and tensions as well (for a brief
review see for example \cite{Perivo} ).

As an alternative to $\Lambda$CDM, DM unified models do not
invoke a cosmological constant. In the framework of general
relativity, non perfect fluids drive the accelerated expansion due to
the negativeness of the viscous pressure, which appears due to the
presence of bulk viscosity. Therefore, a Cold DM (CDM) viscous component
represents a kind of unified DM model that could, in principle, explain the
above mentioned transition of the acceleration without the inclusion of a DE 
component. It is worthy mentioning that measurements of the Hubble constant
show tension with the values obtained from large scale structure
(LSS) and Planck CMB data, which can be alleviated when viscosity is
included in the DM component~\cite{Anand}. The new era of
gravitational waves detector has also opened the possibility to
detect dissipative effects in DM and DE through the dispersion and
dissipation experimented by these waves propagating in a non perfect
fluid. Some constraints on those effects were found
in~\cite{Goswami}.

For neutralino CDM it was pointed out in~\cite
{Hofmann} that a bulk viscosity appears in the CDM
fluid due to the energy transfered from the CDM fluid
to the radiation fluid. From the point of view of cosmological
perturbations, it has been shown that viscous fluid dynamics
provides a simple and accurate framework for extending the
description of cosmological perturbations into the nonlinear regime
~\cite {Blas}. Dissipative DM also appears as a residing component
in a hidden sector, and can reproduce several observational
properties of disk galaxies \cite{Foot_1}, \cite{Foot_2}.
Alternatively, the direct study in astrophysical scenarios, such as
the Bullet Cluster, has been an arena to place constraints on the
DM-DM elastic scattering
cross-section~\cite{Randall},~\cite{Kahlhoefer}. Simulations of this
cluster with the inclusion of self-interacting DM and gas
was performed in~\cite{Robertson}, finding a cross-section of around
$\sigma/m = 2cm^{2}g^{-1}$. Other study presents an indication that
self interaction DM would require a cross-section that varies with
the relative velocity between DM particles in order to
modify the structure of dwarf galaxy dark matter
haloes~\cite{Harvey}. In spite of the fact that the bullet cluster
revealed that the barionic matter has a viscosity much larger than
the DM viscosity, its dissipative negative pressure contribution to 
the accelerated expansion of the universe can be neglected due
to very low density in comparison with the one of the DM.

At background level, where a homogeneous and isotropic space
describes the Universe as a whole, only bulk viscosity is present in
the cosmic fluid and the dissipative pressure must be described by
some relativistic thermodynamical approach to non perfect fluids.
This implies a crucial point in a fully consistent physical
description of the expansion of the Universe using dissipative
processes to generate the transition. Meanwhile in the $\Lambda$CDM
model the acceleration is due to a cosmological constant and the
entropy remains constant, in the case of non perfect fluids it is
necessary to find a solution that not only consistently describes
the kinematics of the Universe, but also that satisfies the
thermodynamical requirements. In the case of a description of
viscous fluids, the Eckart's theory~\cite{Eckart, Eckart2} has been
widely investigated due to its simplicity and became the starting
point to shed some light in the behavior of the dissipative effects
in the late time cosmology~\cite{Avelino2009, Avelino2010, Montiel,
Avelino2013} or in inflationary scenarios~\cite{Padmanabhan}. In the framework of an interacting dark sector, a recent work explores a flat universe with a radiation component and a viscous fluid (DM plus baryons) that interacts with a perfect fluid (DE)~\cite{Almada2020}.  Also a $\Lambda$CDM model with with a dissipative DM, where the viscosity is a  polynomial function of the redshift, has been constrained in~\cite{AlmadaH2020}.

Nevertheless, it is a well known result that the Eckart's theory has
non causal behavior, presenting the problem of superluminal
propagation velocities and some instabilities. So, from the point of
view of a consistent description of the relativistic thermodynamics
of non perfect fluids, it is necessary to include a causal
description such as the one given by the Israel- Stewart (IS)
theory~\cite{Israel, Israel1979, Pavon, Chimento1993, Maartens,
Zimdahl, Maartens1996}.

The aim in this paper is to constraint the respective free
parameters appearing in the recent exact cosmological solutions
found in~\cite{Gonzalez}, for a universe filled only with a
dissipative dark matter component. The constraint was done by using the
Supernova Ia (SNe Ia) and Observational Hubble Data (OHD). These solutions were found in the framework
of the causal thermodynamics described by the IS theory, and are compatible with
a transition between deceleration and acceleration
expansions at background level, within a certain range of the
parameters involved. Since the solution found describes a universe
containing only a dissipative DM as the main component of
the universe, it should only be considered as an adequate
approximation for the late time evolution, where cold DM dominates.
In this sense, these models cannot expected to be fairly
representative of the early time evolution, where ultrarelativistic
matter dominates.

For the solutions was assumed a barotropic EoS for the fluid that
filled the universe, i.e.,
\begin{equation}
p=\left(\gamma-1\right)\rho,
\label{EoS}
\end{equation}
where $p$ is the barotropic pressure, and $\rho$ is the energy density.
These solutions describe the evolution of the universe with
dissipative DM with positive pressure, therefore the
EoS parameter considered lies in the range $1\leq\gamma <2$, where $\gamma=1$
corresponds to a particular solution. Furthermore, the
following Ansatz for the bulk viscosity coefficient, $\xi(\rho)$,
\begin{equation}
\xi (\rho)=\xi_{0}\rho^{s}, \label{xirho}
\end{equation}
was chosen, which has been widely considered as a suitable function between the
bulk viscosity and the energy density of the main fluid. $\xi _{0}$ must be
a positive constant because of the second law of
thermodynamics~\cite{Weinberg1971}. The nonlinear ordinary
differential equation of the IS theory obtained with the above
assumptions has been solved, for example, for different values of
the parameter $s$ in~\cite{Cornejo}; for $s=1/4$ and stiff matter in
~\cite{Harko}. Inflationary solutions were found
in~\cite{Harko1998}. Stability of inflationary solutions were
investigated in~\cite{Chimento1998, Chimento}. For an extensive
review on viscous cosmology in early and late see~\cite{Brevik}.

It is important mentioning that in the formulation of the
thermodynamical approaches of relativistic viscous fluids it is
assumed that the viscous pressure must be lower than the equilibrium
pressure of the fluid ( the near equilibrium condition). Whenever there
are solutions with acceleration at some stage, like, for example,
bulk viscous inflation at early times, or transition between
decelerated and accelerated expansions at late times, the above
condition cannot be fulfilled. Therefore, it is not clearly 
justified the application of the above approach in such situations.

To overcome this issue, deviations from the near equilibrium
condition have been considered within a non linear extension of IS,
as the one proposed in~\cite{Maartens1997}.  Using this
proposal, a nonlinear
extension in accelerated eras occurring at early times, like
inflation or at late times, like phantom behavior, were investigated
in ~\cite{Chimento1} and in~\cite{Cruzphantom}, respectively. The
current accelerated expansion was addressed with a nonlinear
model for viscosity in~\cite{Beesham}. Also, a phase space analysis
of a cosmological model with both viscous radiation and non-viscous
dust was performed in~\cite{Beesham1}, where the viscous pressure
obeys a nonlinear evolution equation. Is important mentioning that 
in ~\cite{Cruz2018} was shown that the inclusion of a cosmological 
constant along with a dissipative DM component allows to 
obey the near equilibrium condition within, in principle, the linear IS theory.

The analytical solution we will analyse in the present article
was obtained using the general expression for the
relaxation time $\tau$~\cite{Maartens1996}, derived from the study
of the causality and stability of the IS theory in~\cite{Hiscock}
\begin{equation}
\frac{\xi}{\left(\rho+p\right)\tau}=c_{b}^{2},
\label{relaxationtime}
\end{equation}
where $c_{b}$ is the speed of bulk viscous perturbations
(non-adiabatic contribution to the speed of sound in a dissipative
fluid without heat flux or shear viscosity). Since the dissipative
speed of sound $V$, is given by $V^{2}= c_{s}^{2}+c_{b}^{2}$, where
$c_{s}^{2}=(\partial p/\partial \rho)_{s}$ is the adiabatic
contribution, then for a barotropic fluid $c_{s}^{2}=\gamma-1$ and
thus $c_{b}^{2}=\epsilon\left(2-\gamma\right)$ with $0<\epsilon\leq
1$, known as the causality condition.  The solution
generalizes the solution found in~\cite{Mathew2017}, where the 
particular expression $\tau=\xi/\rho$ was used, taking besides
the particular values $s=1/2$ and $\gamma=1$.

In a previous work, which included Eq.(\ref{relaxationtime}) for the
relaxation time and a pressureless main fluid, the IS equation was
solved by using an Ansatz for the viscous pressure~\cite{Piattella}.
The conclusion indicates that the full causal theory seems to be
disfavored by the cosmological data. Nevertheless, in the truncated version of the theory,
similar results to those of the $\Lambda CDM$ model were found for a
bulk viscous speed within the interval $10^{-11}\ll c_{b}^{2}\lesssim
10^{-8}$. This last constraint on $c_{b}^{2}$, even though it was
obtained with a suitable Ansatz, and {\bf{it}} does not represent an exact
solution of the theory, teaches us that the non-adiabatic
contribution to the speed of sound must be very small to be
consistent with the cosmological data.

The free parameters of the general analytical solution we will constraint 
against observational data in the present article are $h$,
$\xi_{0}$, $\epsilon$ and $\gamma$ . In the case of one CDM component taking from the beginning, only $h$, $\xi _{0}$ and $\epsilon$ remains free and we find the constraints to obtain a
solution that presents a transition between deceleration to
acceleration expansions. We will also analyse the constraints for the 
case of where all parameters are taken free.

Using the observational constraints obtained for the parameters $h$, $\xi _{0}$ 
and $\epsilon$ for the both cases $\gamma =1$ and $\gamma$ free, we will
discuss the consistence of a fluid description during the cosmic
evolution of the exact solutions representing a dissipative DM 
component. To this aim we evaluate the consistency condition 
$\tau H < 1$ in terms of the constrained parameter values, with $\tau$ 
being the relaxation time and $H$ the Hubble parameter.

This paper is organized as follow: In section II we describe briefly
the causal Israel-Stewart theory and we recall the general
evolution equation for the Hubble parameter $H$.  
We also write down the constraints
for the parameters of the model in order to guaranty a consistent
fluid description. In section III we present the expressions corresponding 
to the analytic solution found in~\cite{Gonzalez} for arbitrary $\gamma$ and
for the dust case, $\gamma=1$, respectively. In section IV we constraint the  
free parameters of the solutions with the Supernovae Ia 
(SNe Ia) and Observational Hubble Data (OHD). In section V we discuss this results comparing them with 
$\Lambda$CDM model and the implication of the parameters's values obtained and their
thermodynamic implications. Finally, in section VI
we present our conclusions taken into account the kinematic properties of 
the solutions, as well as the consistence of a fluid description.

\section{Israel-Stewart formalism}
The model of a dissipative DM component is described by the
Einstein's equations for a flat FRW metric:
\begin{equation}
3H^{2}=\rho, \label{constraint}
\end{equation}
and \begin{equation} 2\dot{H}+3H^{2}=-p-\Pi, \label{acceleration}
\end{equation}
where natural units defined by $8\pi G=c=1$ were used. In addition,
in the IS framework, the transport equation for the viscous pressure
$\Pi $ reads~\cite{Israel1979}
\begin{equation}
\tau\dot{\Pi}+\Pi= -3\xi(\rho) H-\frac{1}{2}\tau\Pi
\left(3H+\frac{\dot{\tau}}{\tau}-\frac{\dot{\xi(\rho)}}{\xi(\rho)}-\frac{\dot{T}}{T}
\right), \label{eqforPi}
\end{equation}
where ``dot" accounts for the derivative with respect to the cosmic
time, $H$ is the Hubble parameter and $T$ is the barotropic
temperature, which takes the form
$T=T_{0}\rho^{\left(\gamma-1\right)/\gamma}$ (Gibbs integrability
condition when $p=\left(\gamma-1\right)\rho$), with $T_{0}$ being a
positive parameter. Using Eqs.(\ref{EoS}), (\ref{xirho}) in
Eq.(\ref{relaxationtime}) we obtain the following expression for the
relaxation time
\begin{equation}
\tau=\frac{\xi_{0}}{\epsilon\gamma\left(2-\gamma\right)}\rho^{s-1}=\frac{3^{s-1}\xi_{0}}{\epsilon\gamma\left(2-\gamma\right)}
H^{2(s-1)} . \label{relaxationtime1}
\end{equation}

In order to obtain a differential equation in terms of the Hubble
parameter, it is neccesary to evaluate the ratios $\dot{\tau}/\tau,
\,\,\dot{\xi}/\xi$ and $\dot{T}/T$, which appear in
Eq.(\ref{eqforPi}). From Eqs.(\ref{constraint}) and
(\ref{acceleration}) the expression for the viscous pressure and its
time derivative can be obtained. Using the above expressions a
nonlinear second order differential equation can be obtained for the
Hubble parameter:

\begin{widetext}
\begin{equation}
\begin{split}
& \ddot{H}+3H\dot{H}+(3)^{1-s}\xi_{0}^{-1}\epsilon\gamma\left(2-\gamma\right)H^{2-2s}\dot{H}-\frac{(2\gamma-1)}{\gamma}H^{-1}\dot{H}^{2}+\frac{9}{4}\gamma\left[1-2\epsilon\left(2-\gamma\right)\right]H^{3} \\
& +\frac{1}{2}(3)^{2-s}\xi_{0}^{-1}\epsilon\gamma^{2}\left(2-\gamma\right)H^{4-2s}=0. \label{eqforH}
\end{split}
\end{equation}
\end{widetext}

We address the reader to see the technical details in ref. \cite{Gonzalez}.
As we shall see in the next section the exact solution was obtained for
the special case $s=1/2$, which allows an important simplification
of Eq. (\ref{eqforH}).  In fact, in this case the simple form
$H\left(t\right)=A\left(t_{s}-t\right)^{-1}$ is a solution of
Eq.(\ref{eqforH}) with a phantom behavior, with $A>0$, $\epsilon=1$
and the restriction $ 0<\gamma<3/2$~\cite{Cruz2017}. Besides, the
solution $H\left(t\right)=A\left(t-t_{s}\right)^{-1}$ can represent
accelerated universes if $A>1$, Milne universes if $A=1$ and
decelerated universes if $A<1$, all of them having an initial
singularity at $t=t_{s}$~\cite{Cruz2017a}.

As it was mentioned in Section I, an important issue that we will
discuss after to constraint the parameters $\xi _{0}$, $\epsilon$,
for the both cases $\gamma =1$ and $\gamma$, is if the found values
satisfy the condition for keeping the fluid description of the
dissipative dark matter component, expressed by the constraint $\tau
H<1$. Using Eq.(\ref{constraint}) for the case $s=1/2$ and
Eq.(\ref{relaxationtime1}), the above inequality leads to the
following constraint between the parameters $\xi _{0}$, $\epsilon$
and $\gamma$
\begin{eqnarray}
\frac{\xi _{0}}{\sqrt{3}\epsilon \gamma(2-\gamma)} \ll 1 .
\label{eq:eq8}
\end{eqnarray}
We will discuss later this condition using the values of $\xi _{0}$,
$\epsilon$, with and without an election of the $\gamma$- value,
obtained from the cosmological data of SNe Ia observations.

\section{The exact cosmological solutions}
Now, we will briefly discuss the two solutions for Eq.(\ref{eqforH})
found in~\cite{Gonzalez} for $s=1/2$ and for the especial cases of
$\gamma\neq 1$ and $\gamma=1$.

\textbf{i)} In the case of $\gamma\neq 1$, the solution for the
Eq.(\ref{eqforH}) can be written as a function of the redshift $z$
as
\begin{equation}
H(z)=C_{3}\left(1+z\right)^{\alpha}\cosh^{\gamma}{\left[\beta\left(\ln{\left(1+z\right)}+C_{4}\right)\right]}, \label{Hgamma}
\end{equation}
where $C_{3}$ and $C_{4}$ are constants given by
\begin{equation}
C_{3}=\frac{H_{0}}{\cosh^{\gamma}{\left(\beta C_{4}\right)}}=H_{0}\left[1-\frac{\left(q_{0}+1-\alpha\right)^{2}}{\gamma^{2}\beta^{2}}\right]^{\frac{\gamma}{2}}, \label{defofC3}
\end{equation}
\begin{equation}
C_{4}=\frac{1}{\beta}\mathop{\mathrm{arctanh}}\left[\frac{\left(q_{0}+1\right)-\alpha}{\gamma\beta}\right], \label{defofC4}
\end{equation}
\begin{equation}
\alpha=\frac{\sqrt{3}\gamma}{2\xi_{0}}\left[\sqrt{3}\xi_{0}+\epsilon\gamma\left(2-\gamma\right)\right], \label{defofalpha}
\end{equation}
\begin{equation}
\beta=\frac{\sqrt{3}}{2\xi_{0}}\sqrt{6\xi_{0}^{2}\epsilon\left(2-\gamma\right)+\epsilon^{2}\gamma^{2}\left(2-\gamma\right)^{2}}. \label{defofbeta}
\end{equation}
In the above expressions $H_{0}$ and $q_{0}$ are the Hubble and deceleration parameters at the present time, respectively, where the deceleration parameter is defined through $q=-1-\dot{H}/H^{2}$. The initial condition $a_{0}=1$ is also used. This solution has a constraint that arises from Eqs.(\ref{defofC3}) and (\ref{defofC4}) that reads
\begin{equation}
\left(\alpha-\gamma\beta\right)-1<q_{0}<\left(\alpha+\gamma\beta\right)-1. \label{constraintq0gamma}
\end{equation}
Since the value of $q_{0}$ will be taken from the observed data, we will check if the above constraints are fulfilled for the values determined for the parameters $\xi_{0}$, $\epsilon$ and $\gamma$ after the constraint of the SNe Ia data.

\textbf{ii)} In the case of $\gamma=1$, the solution of the Eq.(\ref{eqforH}) can be written as
\begin{equation}
H(z)=H_{0}\left[C_{1}\left(1+z\right)^{m_{1}}+C_{2}\left(1+z\right)^{m_{2}}\right], \label{solforH}
\end{equation}
where $H_{0}$ is the Hubble parameter at the present time, and
\begin{equation}
m_{1}=\frac{\sqrt{3}}{2\xi_{0}}\left(\sqrt{3}\xi_{0}+\epsilon+\sqrt{6\xi_{0}^{2}\epsilon+\epsilon^{2}}\right), \label{defofm1}
\end{equation}
\begin{equation}
m_{2}=\frac{\sqrt{3}}{2\xi_{0}}\left(\sqrt{3}\xi_{0}+\epsilon-\sqrt{6\xi_{0}^{2}\epsilon+\epsilon^{2}}\right), \label{defofm2}
\end{equation}
\begin{equation}
C_{1}=\frac{\left(q_{0}+1\right)-m_{2}}{m_{1}-m_{2}}, \label{defofc1}
\end{equation}
\begin{equation}
C_{2}=\frac{m_{1}-\left(q_{0}+1\right)}{m_{1}-m_{2}}. \label{defofc2}
\end{equation}
In the above equations $q_{0}$ is the deceleration parameter at the present time, and the conditions $a_{0}=1$ and $C_{1}+C_{2}=1$ have been set. This solution was previously found and discussed in~\cite{Mathew2017}, but with a particular relation for the relaxation time of the form $\xi_{0}\rho^{s-1}$ (which corresponds to $\alpha=\xi_{0}$ of our Ansatz), instead of the more general relation as Eq.(\ref{relaxationtime1}), which was used in order to obtain the Eq.(\ref{solforH}) in~\cite{Gonzalez}. Even more, this solution has three different behaviors depending on the signs of the constants $C_{1}$ and $C_{2}$. So, for the fit purposes, we limit our analysis to the solution that is most similar to the $\Lambda$CDM model, and that corresponds to the Hubble parameter which fulfills the constraint
\begin{equation}
m_{2}-1<q_{0}<m_{1}-1, \label{constraintHpositive}
\end{equation}
which leads an always positive Hubble parameter.

\section{Constraining the solutions with Supernova Ia and Observational Hubble data sets}
\begin{table*}
\centering
\resizebox{17.94cm}{!}
{
\begin{tabular}{|ll|lllll|lll|}
\hline
\hline
\multicolumn{1}{c}{} & \multicolumn{1}{c@{\hspace{0.5cm}}}{} & \multicolumn{5}{c}{Best fit values} & \multicolumn{1}{c@{\hspace{0.5cm}}}{} & \multicolumn{2}{c}{Goodness of fit}
\\
\cline{3-7}\cline{9-10}
\multicolumn{1}{c}{Data} & \multicolumn{1}{c}{} & \multicolumn{1}{c}{$h$} & \multicolumn{1}{c}{$\Omega_{m}$} & \multicolumn{1}{c}{$x$} & \multicolumn{1}{c}{$\epsilon$} & \multicolumn{1}{c}{$\gamma$} & \multicolumn{1}{c}{} & \multicolumn{1}{c}{$\chi^{2}_{min}$} & \multicolumn{1}{c}{$BIC$} 
\\
\hline
\multicolumn{10}{c}{$\Lambda$CDM model}
\\
\multicolumn{1}{c}{SNe Ia} & \multicolumn{1}{c}{} & \multicolumn{1}{c}{$0.732_{-0.018}^{+0.017}$} & \multicolumn{1}{c}{$0.299_{-0.022}^{+0.022}$} &  \multicolumn{1}{c}{$\cdots$} & \multicolumn{1}{c}{$\cdots$} & \multicolumn{1}{c}{$\cdots$} & \multicolumn{1}{c}{} & \multicolumn{1}{c}{1026.9} & \multicolumn{1}{c}{1040.8}
\\
\multicolumn{1}{c}{OHD} & \multicolumn{1}{c}{} &  \multicolumn{1}{c}{$0.715_{-0.010}^{+0.010}$} & \multicolumn{1}{c}{$0.248_{-0.014}^{+0.015}$} & \multicolumn{1}{c}{$\cdots$} & \multicolumn{1}{c}{$\cdots$} & \multicolumn{1}{c}{$\cdots$} & \multicolumn{1}{c}{} & \multicolumn{1}{c}{27.9} & \multicolumn{1}{c}{35.7}
\\
\multicolumn{1}{c}{SNe Ia + OHD} & \multicolumn{1}{c}{} &  \multicolumn{1}{c}{$0.705_{-0.009}^{+0.009}$} & \multicolumn{1}{c}{$0.265_{-0.012}^{+0.013}$} & \multicolumn{1}{c}{$\cdots$} & \multicolumn{1}{c}{$\cdots$} & \multicolumn{1}{c}{$\cdots$} & \multicolumn{1}{c}{} & \multicolumn{1}{c}{1057.1} & \multicolumn{1}{c}{1071.1}
\\
\hline
\multicolumn{10}{c}{Exact cosmological solution with $\gamma\neq 1$}
\\
\multicolumn{1}{c}{SNe Ia} & \multicolumn{1}{c}{} & \multicolumn{1}{c}{$0.732_{-0.017}^{+0.017}$} & \multicolumn{1}{c}{$1$} & \multicolumn{1}{c}{$1.288_{-0.276}^{+0.199}$} & \multicolumn{1}{c}{$0.709_{-0.143}^{+0.181}$} & \multicolumn{1}{c}{$1.194_{-0.139}^{+0.177}$} & \multicolumn{1}{c}{} & \multicolumn{1}{c}{$1030.0$} & \multicolumn{1}{c}{$1057.8$}
\\
\multicolumn{1}{c}{OHD} & \multicolumn{1}{c}{} & \multicolumn{1}{c}{$0.735_{-0.007}^{+0.007}$} & \multicolumn{1}{c}{$1$} & \multicolumn{1}{c}{$1.422_{-0.192}^{+0.108}$} & \multicolumn{1}{c}{$0.445_{-0.056}^{+0.177}$} & \multicolumn{1}{c}{$1.108_{-0.082}^{+0.213}$} & \multicolumn{1}{c}{} & \multicolumn{1}{c}{$62.2$} & \multicolumn{1}{c}{77.9}
\\
\multicolumn{1}{c}{SNe Ia + OHD} & \multicolumn{1}{c}{} & \multicolumn{1}{c}{$0.731_{-0.006}^{+0.006}$} & \multicolumn{1}{c}{$1$} & \multicolumn{1}{c}{$1.488_{-0.125}^{+0.061}$} & \multicolumn{1}{c}{$0.396_{-0.022}^{+0.045}$} & \multicolumn{1}{c}{$1.044_{-0.033}^{+0.078}$} & \multicolumn{1}{c}{} & \multicolumn{1}{c}{$1089.2$} & \multicolumn{1}{c}{$1117.2$}
\\
\hline
\multicolumn{10}{c}{Exact cosmological solution with $\gamma=1$}
\\
\multicolumn{1}{c}{SNe Ia} & \multicolumn{1}{c}{} & \multicolumn{1}{c}{$0.732_{-0.017}^{+0.017}$} & \multicolumn{1}{c}{$1$} & \multicolumn{1}{c}{$1.161_{-0.314}^{+0.282}$} & \multicolumn{1}{c}{$0.553_{-0.068}^{+0.126}$} & \multicolumn{1}{c}{$1$} & \multicolumn{1}{c}{} & \multicolumn{1}{c}{$1027.2$} & \multicolumn{1}{c}{$1048.1$}
\\
\multicolumn{1}{c}{OHD} & \multicolumn{1}{c}{} & \multicolumn{1}{c}{$0.733_{-0.006}^{+0.006}$} & \multicolumn{1}{c}{$1$} & \multicolumn{1}{c}{$1.408_{-0.203}^{+0.119}$} & \multicolumn{1}{c}{$0.378_{-0.016}^{+0.032}$} & \multicolumn{1}{c}{$1$} & \multicolumn{1}{c}{} & \multicolumn{1}{c}{$39.7$} & \multicolumn{1}{c}{$51.5$}
\\
\multicolumn{1}{c}{SNe Ia + OHD} & \multicolumn{1}{c}{} &  \multicolumn{1}{c}{$0.730_{-0.006}^{+0.006}$} & \multicolumn{1}{c}{$1$} & \multicolumn{1}{c}{$1.479_{-0.132}^{+0.068}$} & \multicolumn{1}{c}{$0.371_{-0.009}^{+0.018}$} & \multicolumn{1}{c}{$1$} & \multicolumn{1}{c}{} & \multicolumn{1}{c}{$1083.0$} & \multicolumn{1}{c}{$1104.0$}
\\
\hline\hline
\end{tabular}
}
\caption{Best-fit values for each model free parameters $\theta$, as well as the respective goodness of fit criteria, obtained in the MCMC analysis. The first row shows the best-fit values for the standard cosmological model $\Lambda$CDM, while the second and third rows shows the best-fit parameters of the exact cosmological solution with $\gamma\neq 1$ and $\gamma =1$, respectively. The uncertainties correspond to $1\sigma$ ($68.3\%$) of confidence level (CL). The best-fits values for the $\Lambda$CDM model are used for the sake of comparison with the exact cosmological solutions.} \label{bestfittable}
\end{table*}

\begin{figure*}
\includegraphics[scale=0.5]{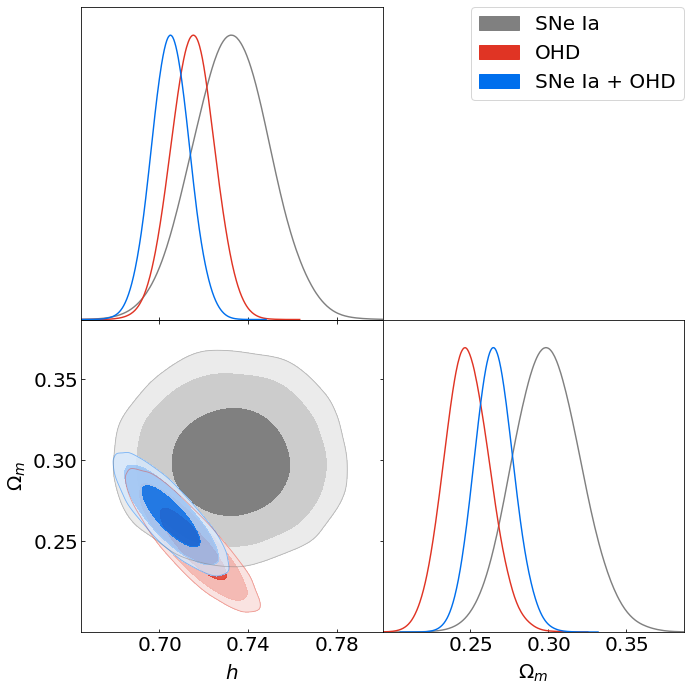}
\caption{Joint and marginalized constraint of $h$ and $\Omega_{m}$ obtained in the MCMC analysis for the $\Lambda CDM$ model. The admissible regions correspond to $1\sigma\left(68.3\%\right)$, $2\sigma\left(95.5\%\right)$, and $3\sigma\left(99.7\%\right)$ confidence level (CL), respectively. The best-fit values for each parameter are shown in Table \ref{bestfittable}.}
\label{triangleLCDM}
\end{figure*}

\begin{figure*}
\includegraphics[scale=0.5]{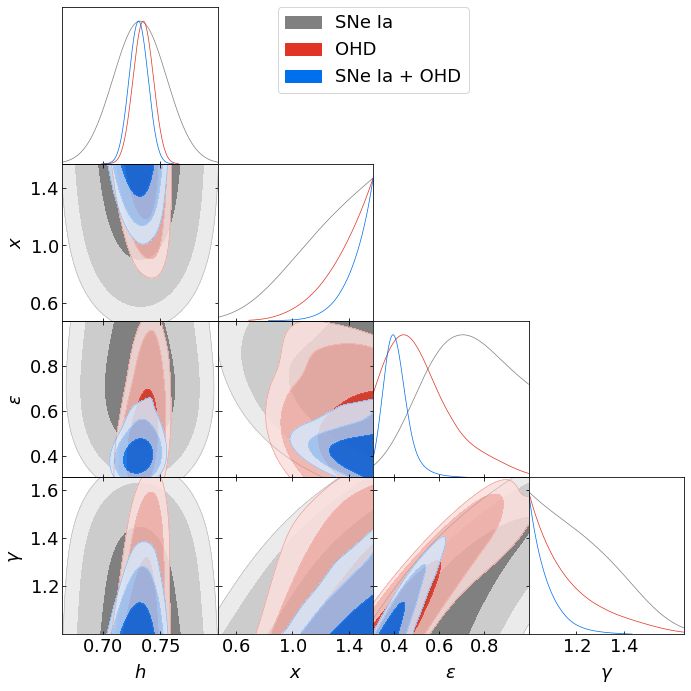}
\caption{Joint and marginalized constraint of $h$, $x$, $\epsilon$ and $\gamma$ obtained in the MCMC analysis for the exact cosmological solution with $\gamma\neq 1$. The admissible regions correspond to $1\sigma\left(68.3\%\right)$,
$2\sigma\left(95.5\%\right)$, and $3\sigma\left(99.7\%\right)$ confidence level (CL), respectively. The best-fit values for each parameter are shown in Table \ref{bestfittable}.}
\label{trianglegeneral}
\end{figure*}

\begin{figure*}
\includegraphics[scale=0.5]{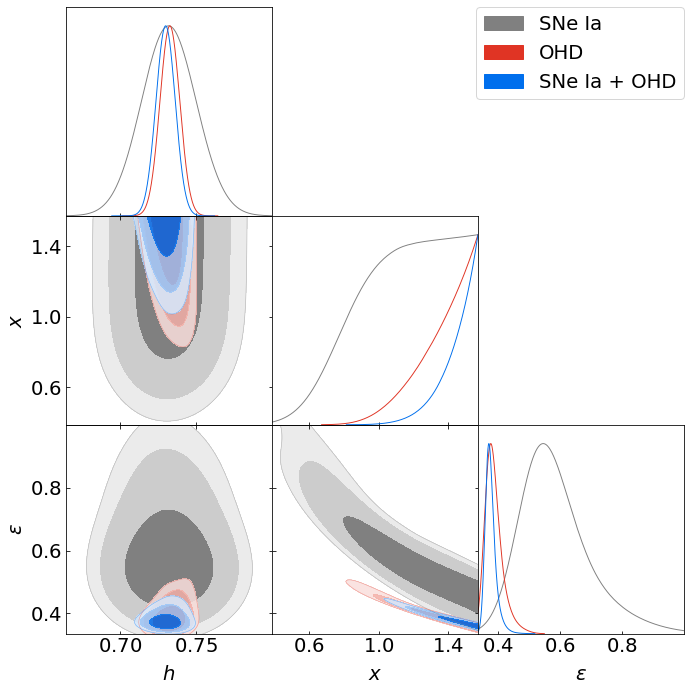}
\caption{Joint and marginalized constraint of $h$, $x$ and $\epsilon$ obtained in the MCMC analysis for the exact cosmological solution with $\gamma=1$. The admissible regions correspond to $1\sigma\left(68.3\%\right)$, $2\sigma\left(95.5\%\right)$, and $3\sigma\left(99.7\%\right)$ confidence level (CL), respectively. The best-fit values for each parameter are shown in Table
\ref{bestfittable}.} \label{triangleparticular}
\end{figure*}

We shall constrain the free parameters of the solutions presented in the above section with the Supernova Ia data (SNe Ia) and the Observational Hubble Data (OHD). To do so, we compute the best-fit parameters with the affine-invariant Markov Chain Monte Carlo Method (MCMC)~\cite{Goodman}, implemented in the pure-Python code \textit{emcee}~\cite{Foreman}, by setting 50 chains or ``walkers" with 10000 steps and 10000 burn-in steps; this last ones in order to let the walkers explore the parameters space and get settled in the maximum of the probability density. As a likelihood function we use the following Gaussian distribution
\begin{equation}
\mathcal{L}\propto\exp{\left(-\frac{\chi_{I}^{2}}{2}\right)}, \label{distribution}
\end{equation}
where $\chi_{I}^{2}$ is the merit function with $I$ representing each data set, namely SNe Ia, OHD and their joint analysis $\chi_{joint}^{2}=\chi_{SNe}^{2}+\chi_{OHD}^{2}$. Therefore, to impose the constraint, we use the Pantheon SNe Ia sample~\cite{Scolnic} and the compilation of OHD provided by Magaña \textit{et al.}~\cite{Magana}. 

In the first one, the sample consist in 1048 data points in the redshift range $0.01\leq z\leq 2.3$, that is a compilation of 279 SNe Ia data discovered by the Pan-STARRS1 (PS1) Medium Deep Survey, combined with the distance estimates of SNe Ia from the Sloan Digital Sky Survey (SDSS), Supernova Legacy Survey (SNLS), and various low-z and Hubble Space Telescope (HST) samples, where the distance estimator is obtained using a modified version of the Tripp formula~\cite{Tripp} with two nuisance parameters calibrated to zero with the method ``BEAMS with Bias Correction'' (BBC) proposed by Kessler and Scolnic~\cite{Kessler}. Hence, the observational distance modulus for each SNe Ia at a certain redshift $z_{i}$ is given by the formula
\begin{equation}
\mu_{i} = m_{B,i}-\mathcal{M} \label{mu},
\end{equation}
where $m_{B,i}$ is the apparent B-band magnitude of a fiducial SNe Ia and $\mathcal{M}$ is a nuisance parameter. Considering that the Pantheon sample give directly the corrected apparent magnitude for each SNe Ia, the merit function for the SNe Ia data sample can be constructed in matrix notation as
\begin{equation}
\chi_{SNe}^{2}=\mathbf{M}^{\dagger}\mathbf{C^{-1}}\mathbf{M} \label{MeritSNe},
\end{equation}
where $\mathbf{M}=\mathbf{m}_{B}-\boldsymbol\mu_{th}-\boldsymbol{\mathcal{M}}$ and $\mathbf{C}$ is the total covariance matrix, given by
\begin{equation}
\mathbf{C}=\mathbf{D}_{stat}+\mathbf{C}_{sys} \label{CovarianceMatrix},
\end{equation}
where the diagonal matrix $\mathbf{D}_{stat}$ denotes the statistical uncertainties obtained by adding in quadrature the uncertainties of $m_{B}$ for each redshift, associated with the BBC method, while $\mathbf{C}_{sys}$ denotes the systematic uncertainties in the BBC approach. On the other hand, the theoretical distance modulus for each SNe Ia at a certain redshift $z_{i}$ in a flat FLRW spacetime for a given model is defined through the relation

\begin{equation}
\mu_{th}\left(z_{i},\theta\right)=5\log_{10}{\left[\frac{d_{L}\left(z_{i},\theta\right)}{Mpc}\right]}+\bar{\mu}, \label{mutheoretical}
\end{equation}
where $\theta$ encompasses the free parameters of the respective solution, $\bar{\mu}=5\log_{10}{(c)}+25$ with $c$ the speed of light and $d_{L}$ is the luminosity distance which takes the form
\begin{equation}
d_{L}(z_{i},\theta)=(1+z_{i})\int_{0}^{z_{i}}{\frac{dz'}{H(z',\theta)}} \label{luminosity}.
\end{equation}
In order to reduce the number of free parameters and marginalized over the nuisance parameter $\mathcal{M}$, we define $\bar{\mathcal{M}}=\bar{\mu}+\mathcal{M}$ and the merit function (\ref{MeritSNe}) can be expanded as~\cite{Lazkoz}
\begin{equation}
\chi^{2}_{SNe}=A(z,\theta)-2B(z,\theta)\bar{\mathcal{M}}+C(z,\theta)\bar{\mathcal{M}}^{2} \label{expandedmerit},
\end{equation}
where
\begin{equation}
A(z,\theta)=\mathbf{M}(z,\theta,\bar{\mathcal{M}}=0)^{\dagger}\mathbf{C^{-1}}\mathbf{M}(z,\theta,\bar{\mathcal{M}}=0),
\end{equation}
\begin{equation}
B(z,\theta)=\mathbf{M}(z,\theta,\bar{\mathcal{M}}=0)^{\dagger}\mathbf{C^{-1}}\mathbf{1},
\end{equation}
\begin{equation}
C(z,\theta)=\mathbf{1}^{\dagger}\mathbf{C^{-1}}\mathbf{1}.
\end{equation}
Hence, minimizing the expression (\ref{expandedmerit}) whit respect to $\bar{\mathcal{M}}$ gives $\bar{\mathcal{M}}=B/C$ and the merit function reduces to
\begin{equation}
\chi_{SNe}^{2}\Bigr\rvert_{min}=A(z,\theta)-\frac{B(z,\theta)^{2}}{C} \label{MeritSNefinal},
\end{equation}
that clearly only depends on the free parameters of the respective solution. It is important to note that the merit function given by (\ref{MeritSNe}) provides the same information
as the function given by (\ref{MeritSNefinal}); this is because the best fit parameters minimize the merit function. Then, $\chi^{2}_{min}$ gives an indication of the goodness of fit: the smaller its value, the better is the fit.

In the second one, the sample consist in 51 data points in the redshift range $0.07\leq z\leq 2.36$, where 31 data points are obtained by the Differential Age (DA) method~\cite{Jimenez} which implies that these data points are model independent. The remaining 20 points come from BAO measurements, assuming that the $H(z)$ data obtained come from independent measurements. Hence, the merit function for the OHD data sample can be constructed as
\begin{equation}
\chi^{2}_{OHD}=\sum_{i=1}^{51}{\left[\frac{H_{i}-H_{th}(z_{i},\theta)}{\sigma_{H,i}}\right]^{2}} \label{MeritOHD},
\end{equation}
where $H_{i}$ is the observational Hubble parameter at redshift $z_{i}$ with associated error $\sigma_{H,i}$, provided by the OHD sample considered, $H_{th}$ is the theoretical Hubble parameter at the same redshift, provided by the solutions, and $\theta$ encompasses the free parameters of the respective solution.

The two cosmological solutions are contrasted with the SNe Ia and OHD data through their corresponding Hubble parameter (\ref{Hgamma}) and (\ref{solforH}). Because the solutions correspond to only matter as a dominant component, we have to impose $\Omega_{m}=1$. So, for the solution with $\gamma\neq 1$ their free parameters are $\theta=\{H_{0},\xi_{0},\epsilon,\gamma\}$ and the solution with $\gamma = 1$ are $\theta=\{H_{0},\xi_{0},\epsilon\}$. Even more, dimensionless parameters for the fit are required, where $\epsilon$ and $\gamma$ are already dimensionless. So, we replace $H_{0}$ for the dimensionless Hubble parameter $h$, where
\begin{equation}
H_{0}=100\;km\;s^{-1}\;Mpc^{-1}\times h \label{H0dimensionless},
\end{equation}
and a dimensionless $\xi_{0}$ required the following redefinition
\begin{equation}
\xi_{0}\rightarrow H_{0}^{1-2s}\xi_{0}, \label{xi0dimensionless}
\end{equation}
where, considering that the solutions are obtained for $s=1/2$, then $\xi_{0}$ it is in particular also dimensionless. In consequence, for $h$ we use a Gaussian prior according to the value reporting by A. G. Riess \textit{et al.} in~\cite{Riess2016} of $H_{0}=73.24\pm 1.74 \;km\;s^{-1}\;Mpc^{-1}$ wich is measured with a $2.4\%$ of uncertainty, for $\epsilon$ and $\gamma$ we use the flat priors $0<\epsilon<1$ and $1<\gamma<2$, and for $\xi_{0}$ we make the change of variable $\xi_{0}=\xi_{0}(x)=\tan{(x)}$ for which we use the flat prior $0<x<\pi/2$; this last one in order to simplify the sampling of the full parameter space using in the MCMC analysis. It is important to mention that in both cases we use the actual value of the deceleration parameter, $q_{0}=-0.6$, as initial condition~\cite{Planck}. In the solution with $\gamma\neq 1$ we need to use as a prior the restriction given by Eq.(\ref{constraintq0gamma}), in order to avoid a complex Hubble parameter during the fit; and in the solution with $\gamma=1$ we need to use as a prior the restriction given by Eq.(\ref{constraintHpositive}), in order to obtain a positive Hubble parameter. Moreover, the $a$ parameter in the \textit{emcee} code is modified in order to obtain a mean acceptance fraction between $0.2$ and $0.5$~\cite{Foreman}.

\section{Results and discussion}
Both solutions will be compared with the standard cosmological model $\Lambda$CDM, whose
respective Hubble parameter as a function of the redshift is given by
\begin{equation}
H_{\Lambda CDM}(z)=H_{0}\sqrt{1-\Omega_{m}+\Omega_{m}\left(1+z\right)^{3}}, \label{LambdaCDM}
\end{equation}
with their respective free parameters $\theta=\{h,\Omega_{m}\}$, where for $\Omega_{m}$ we use the flat prior $0<\Omega_{m}<1$, and for $h$ the we use the same Gaussian prior as for the exact cosmological solutions. Even more, in order to compare the goodness of the fits statistically, we will use the Bayesian Information Criterion (BIC)~\cite{Schwarz}, defined as
\begin{equation}
BIC=\theta_{N}\ln{\left(n\right)}-2\ln{\left(\mathcal{L}_{max}\right)}, \label{BIC}
\end{equation}
where $\mathcal{L}_{max}$ is the maximum value of the likelihood function, calculated for the best-fit parameters, $\theta_{N}$ the total number of free parameters of the model and $n$ is the total number of the respective data sample. This criteria tries to solve the problem of maximizing the likelihood function by adding free parameters resulting in over-fitting. To do so, the criteria introduces a penalization that depends on both, the total number of free parameters of each model and the total observational data. The model statistically favored by observations, as compared to the others, corresponds to the one with the smallest value of BIC, where a difference of $2-6$ in BIC between two models is considered as evidence against the model with the higher BIC, a difference of $6-10$ in BIC is already strong evidence, and a difference $>10$ in BIC represents definitely a very strong evidence.

\begin{table}[h]
\centering
\resizebox{7cm}{!}
{
\begin{tabular}{ccc}
\hline\hline
Data & \multicolumn{1}{c@{\hspace{0.5cm}}}{} & $\xi_{0}$ 
\\
\hline
\multicolumn{3}{c}{Exact cosmological solution with $\gamma\neq 1$}
\\
SNe Ia & & $3.441_{-1.842}^{+8.464}$
\\
OHD & & $6.671_{-3.851}^{+17.828}$
\\
SNe Ia + OHD & & $12.050_{-7.307}^{+33.822}$ 
\\
\hline
\multicolumn{3}{c}{Exact cosmological solution with $\gamma= 1$}
\\
SNe Ia & & $2.302_{-1.171}^{+5.480}$
\\
OHD & & $6.088_{-3.478}^{+16.730}$
\\
SNe Ia + OHD & & $10.863_{-6.470}^{+31.152}$
\\
\hline\hline
\end{tabular}
}
\caption{Best-fit values for the free parameter $\xi_{0}$ obtained from the best-fit values of $x$, indicated in the Table \ref{bestfittable}, and the relation $\xi_{0}=tan{(x)}$.} \label{xi0table}
\end{table}

The best-fit values for the $\Lambda$CDM model and the exact cosmological solutions, as well as the goodness of fit criteria are shown in Table \ref{bestfittable}. In Figs.\ref{triangleLCDM}-\ref{triangleparticular} we depict their respective joint credible regions for combinations of their respective free parameters. From them, we are be able to conclude that:

\begin{itemize}
\item[1.-] The $\Lambda$CDM model has the lower values of $\chi^{2}_{min}$ and BIC, i. e., it is the model more favored by the observations. Focusing in the values of $\chi^{2}_{min}$, the solution with $\gamma=1$ is as suited to describe the SNe Ia data as the $\Lambda$CDM model does, with a difference in $\chi^{2}_{min}$ smaller than $0.3$. But, from the joint credible regions graphics it is possible to see that the SNe Ia data constricts less the free parameters than the OHD data and the joint data analysis. On the other hand, focusing in the BIC criteria, the smallest BIC difference occurs between the $\Lambda$CDM model and the solution with $\gamma=1$ reaching already for this difference the value 7.3 for the SNe Ia data. This has the consequence that the other solutions for $\gamma \neq 1$ are disfavored by the data. Moreover, the observations favor models where the recent acceleration expansion of the Universe is due to DE, instead of the models where the acceleration is due to the dissipative effects that experiments the DM. Even more, the exact cosmological solution with $\gamma=1$ has lower values of $\chi^{2}_{min}$ and BIC than the solutions with $\gamma\neq 1$, i. e., the observations favor the solutions where a CDM is considered.

\item[2.-] The main issue of the solutions arise from the best-fit values obtained for $\epsilon$, which clearly are inconsistent with the value of $10^{-11}\ll\epsilon\lesssim 10^{-8}$ reported in~\cite{Piattella}, in order to be consistent with the properties of structure formation. 

\item[3.-] In order to fulfill the condition $\tau H\ll 1$, given by the Eq.(\ref{eq:eq8}), it is necessary, in the best scenario, that $\xi_{0}\ll 2\sqrt{3}$. From the values of $\xi_{0}$ shown in the Table \ref{xi0table}, it is possible to see that the value obtained from the SNe Ia data for both solutions and for the lower interval, gives a value close to $\sqrt{3}$, and for OHD data a value close to $2\sqrt{3}$, which are clearly greater than $2\sqrt{3}$ for the joint data analysis. Therefore, the condition $\tau H\ll 1$ is not fulfilled by the exact cosmological solution any of both cases, nevertheless, there is the possibility that the fluid condition can be fulfilled in some regime, improving the fit data, or under new considerations when studying the proposed cosmological model. So, this claim is not conclusive. 

\item[4.-] In natural units $\xi_{0}$ is a dimensionless parameter.  In terms of physical units, it has no viscosity units due to the form in which it was defined. Nevertheless, it is possible to evaluate the dissipative pressure, for example, at the present time, in order to get an estimation of the size of the values involved.  For the present time we obtained that $\Pi\approx 10^{-20}\,Pa$, which is a very low pressure, in comparison, for example, with the values obtained in the Eckart's framework (see~\cite{Brevik11}).

\item[5.-] The possible explanation for the principal drawbacks presented by the exact cosmological solution for $\gamma\neq1$ and $\gamma=1$ could be related to the particular election for the bulk viscosity coefficient (see Eq.(\ref{xirho})), which is in this case proportional to the root of the DM density and it is the responsible of the recent acceleration expansion of the universe in this model. Because $\rho\rightarrow\infty$ when $z\rightarrow\infty$, and $\rho\rightarrow 0$ when $z\rightarrow -1$, the bulk viscosity becomes relevant in the past and negligible in the future, which is when the Universe experiments the acceleration in its expansion. Therefore, in order that the bulk viscosity becomes relevant at present and future time, it is necessary to increase the value of $\xi_{0}$, which inevitably prevents to fulfil the near equilibrium condition $\tau H\ll 1$, alternatively, the rise of the $\epsilon$ value would be required. This fact can be observed in the Figs. \ref{trianglegeneral} and \ref{triangleparticular}, (most clearly in the Fig. \ref{triangleparticular}), where for a lower values of $\xi_{0}$ larger $\epsilon$ values are obtained and vice versa. It is worthwhile mentioning that, because $\epsilon$ cannot be larger than one, $\xi_{0}$ has a non-zero lower bound; and for this minimum value, $\xi_{0}\rightarrow\infty$.
\end{itemize}

\section{Conclusions}
We have tested a
cosmological model described by an analytical solution recently found 
in~\cite{Gonzalez} for $s=1/2$ and
for arbitrary $\gamma$, including the particular case when $\gamma=1$,
by constraining it against Supernovae Ia and Observational Hubble Data.
The solution gives the time evolution of the Hubble
parameter in the framework of the full causal thermodynamics of
Israel-Stewart. This solution was obtained considering a bulk
viscous coefficient with the dependence $\xi=\xi_{0}\rho^{1/2}$, the 
general expression given by Eq.(\ref{relaxationtime}) for the
relaxation time, and for a fluid with a barotropic EoS
$p=\left(\gamma-1\right)\rho$. The results of the constraints still
indicate that the $\Lambda CDM$ model is statistically the most
favored model by the observations.

The lesson that we have
learned here is that unified DM models succeed to display the
transition between decelerated and accelerated expansions, which is
an essential a feature supported by the observational data, without
invoking a cosmological constant or some other form of dark energy.
Nevertheless, as it was found in~\cite{Piattella}, only a very small
value of $\epsilon$ is consistent with the structure formation, while
the numerical value we found from the best fit to the data leads to
inconsistencies with the values required at perturbative level. 

It is relevant to mention that the exact solution constrained in this paper displays naturally, for some parameter values,  the transition between decelerated and accelerated expansions. Other solutions found in the literature within the IS framework and for the same election for the bulk viscosity coefficient, (see for instance \cite{Cruzpowerlaw}), which are described by the power law behavior $H\left( t\right) =A\left( t_{s}-t\right) ^{-1}$, do not display the same natural transition. In fact,  depending on the  parameter $A$, they represent monotonically accelerated or decelerated solutions. For the case of an accelerated expansion a large non adiabatic contribution to the speed of sound is required. These two investigated solutions have in common the election of a bulk viscosity coefficient, which grows with the energy density of DM. This Ansatz has been made due to the simplicity the master equation acquires within the IS formalism. Nevertheless, from the physical point of view, this choice implies that the negative dissipative pressure grows with the redshift, while the inverse behavior leads to an accelerated late time expansion. 

 The above results indicate that in the framework of the causal thermodynamics theory of dissipative fluids, accelerated solutions can in fact be obtained, nevertheless the non adiabatic contribution to the speed of sound happen to be large, in contradiction with the conclusions of perturbation analysis. This result can be inferred when the general expression for the relaxation time $\tau$, given by Eq.(\ref{relaxationtime}), is used. In some previous results, like the one displayed in ~\cite{Mathew2017}, where $\epsilon$ is set equal to one from the beginning, the consequences of this drawback was not properly acknowledged. This result is also consistent with the mathematical condition found in~\cite{Gonzalez}, where the exact solution displays an accelerated expansion only if $\epsilon>1/18$.  

 Moreover,  we have shown that the values of the parameters found from the data constraints lead to an inconsistency of the fluid description of the dissipative dark matter component. In fact, the best fit parameters indicate that the required condition $\tau H\ll 1$ cannot be fulfilled by the solution. This result is consistent with the the basic assumption in the thermodynamic approaches of relativistic viscous fluids, which asserts that the viscous stress should be lower than the equilibrium pressure of the fluid. This is the so-called near equilibrium condition. When the negative pressure comes only from the dissipation, the above condition is not fulfilled. The condition $\tau H\ll 1$ means that particles of the fluid has an interaction rate that allows to keep the thermal equilibrium, adjusting more rapidly that the natural time-scale defined by the expansion time $H^{-1}$~\cite{Maartens1996}. Therefore, it is expected that the condition $\tau H\ll 1$ will not be fulfilled by the parameters of an exact solution when this describes accelerated expansions. Nevertheless, as it was previously mentioned, this feature does not rule out the possibility that this condition be fulfilled in some other region of the the solution.

 Extensions of the IS approach, which consider non-linear effects allow deviations from the equilibrium. This could represent a possible solution to the technical difficulties just mentioned above, and to some extent one scenario of it has been explored in ~\cite{Cruz2017}, for phantom-type solutions. We expect to go further and extend the analytic solution including this nonlinear generalization elsewhere. 

\begin{acknowledgments}
We thank Arturo Avelino for useful discussions. This article was partially supported by Dicyt from Universidad de Santiago de Chile, through Grants $N^{\circ}$ $041831PA$ (G.P.) and $N^{\circ}$ $041831CM$ (N.C.). E.G. was supported by Proyecto POSTDOC\_DICYT, c\'odigo 041931CM\_POSTDOC, Universidad de Santiago de Chile and partially supported by CONICYT-PCHA/Doctorado Nacional/2016-21160331.
\end{acknowledgments}


\begin{thebibliography}{99}

\bibitem{Riess} A. G. Riess \textit{et al.} (Supernova Search Team), Astron. J. \textbf{116}, 1009 (1998).

\bibitem{Perlmutter} S. Perlmutter \textit{et al.} (Supernova Cosmology Project), Astrophys. J. \textbf{517}, 565 (1999).

\bibitem{WMAP} E. Komatsu \textit{et al.} (WMAP Collaboration), Astrophys. J. Suppl. Ser. \textbf{192}, 18 (2011).

\bibitem{Planck} P. A. R. Ade \textit{et al.} (Planck Collaboration), Astron. Astrophys. \textbf{571}, A16 (2014).

\bibitem{Planck2016} P. A. R. Ade \textit{et al.} (Planck Collaboration), Astron. Astrophys. \textbf{594}, A13 (2016).

\bibitem{Moresco} M. Moresco, \textit{et al.}, J. Cosmol. Astropart. Phys. \textbf{05}, 014 (2016).

\bibitem{Weinberg} S. Weinberg, Rev. Mod. Phys. \textbf{61}, 1 (1989).

\bibitem{Carroll} S. M. Carroll, W. H. Press and  E. L. Turner, Ann. Rev. Astron. Astrophys. \textbf{30}, 499 (1992).

\bibitem{Turner} M. S. Turner, Phys. Rep. \textbf{333}, 619 (2000).

\bibitem{Sahni} V. Sahni and A. Starobinsky, Int. J. Mod. Phys. D \textbf{9}, 373 (2000).

\bibitem{Carroll2001} S. M. Carroll, Living Rev. Rel. \textbf{4}, 1 (2001).

\bibitem{Padmanabhan2003} T. Padmanabhan, Phys. Rep. \textbf{380}, 335 (2003).

\bibitem{Peebles} P. J. E. Peebles and B. Ratra, Rev. Mod. Phys. \textbf{75}, 559 (2003).

\bibitem{Steinhardt} P. J. Steinhardt, in Critical Problems in Physics, edited by V.L. Fitch and D.R. Marlow (Princeton University, Princeton, NJ, 1997).

\bibitem{Zlatev} I. Zlatev, L. Wang, and P. J. Steinhardt, Phys. Rev. Lett. \textbf{82}, 896 (1999).

\bibitem{Perivo} L. Perivolaropoulos, J. Cosmol. \textbf{15}, 6054 (2011).

\bibitem{Anand} S. Anand, P. Chaubal, A. Mazumdar, and S. Mohanty, J. Cosmol. Astropart. Phys. \textbf{11}, 005 (2017).

\bibitem{Goswami} G. Goswami, G. K. Chakravarty, S. Mohanty, and A. R. Prasanna, Phys. Rev. D \textbf{95}, 103509 (2017).

\bibitem {Hofmann} S. Hofmann, D. J. Schwarz, and H. Stöcker, Phys. Rev. D 64,
083507 (2001).

\bibitem {Blas} D. Blas, S. Floerchinger, M. Garny, N. Tatradis and U. A. Wiedemann,
JCAP 1511 (2015) 049

\bibitem{Foot_1} R. Foot and  S. Vagnozzi, Phys. Rev. D \textbf{91} (2015) 023512.

\bibitem{Foot_2} R. Foot and S. Vagnozzi, JCAP 1607 (2016) , 07, 013.

\bibitem{Randall} S. W. Randall, M. Markevitch, D. Clowe, A. H. Gonzalez and M. Brada$\check{a}$, 2008, ApJ, 679, 1173

\bibitem{Kahlhoefer} F. Kahlhoefer, K. Schmidt-Hoberg, J. Kummer and S.Sarkar, 2015, MNRAS,
452, L54

\bibitem{Robertson} A. Robertson, R. Massey and V. Eke, Mon.Not.Roy.Astron.Soc. 465 (2017) no.1, 569-587

\bibitem{Harvey} D. Harvey, A. Robertson, R. Massey and I. G. McCarthy, Mon.Not.Roy.Astron.Soc. 488 (2019)
no.2, 1572-1579

\bibitem{Eckart} C. Eckart, Phys. Rev. \textbf{58}, 267 (1940).

\bibitem{Eckart2} C. Eckart, Phys. Rev. \textbf{58},  919 (1940).

\bibitem{Avelino2009} A. Avelino and U. Nucamendi, J. Cosmol. Astropart. Phys. \textbf{04}, 006 (2009).

\bibitem{Avelino2010} A. Avelino and U. Nucamendi, J. Cosmol. Astropart. Phys. \textbf{08}, 009 (2010).

\bibitem{Montiel} A. Montiel and N. Bret\'on, J. Cosmol. Astropart. Phys. \textbf{08}, 023 (2011).

\bibitem{Avelino2013} A. Avelino, Y. Leyva, and L. A. Ure\~na-L\'opez, Phys. Rev. D \textbf{88}, 123004 (2013).

\bibitem{Padmanabhan} T. Padmanabhan and S. M. Chitre, Phys. Lett. A \textbf{120}, 433 (1987).

\bibitem{Almada2020} A. Hernández-Almada, M. A. García-Aspeitia, J. Magaña,and V. Motta,  Phys. Rev. D \textbf{101} (2020) 063516.

\bibitem{AlmadaH2020} L. Herrera-Zamorano, A. Hernández-Almada, and M.A. García-Aspeitia, Eur. Phys. J. C \textbf{80}(2020) 637. 

\bibitem{Israel} W. Israel, Ann. Phys. (N.Y.) \textbf{100}, 310 (1976).

\bibitem{Israel1979} W. Israel and J. M. Stewart, Ann. Phys. (N.Y.) {\textbf{118}}, 341 (1979).

\bibitem{Pavon} D. Pav\'on, Class. Quant. Grav. \textbf{7}, 487 (1990).

\bibitem{Chimento1993} L. P. Chimento and A. S. Jakubi, Class. Quant. Grav. \textbf{10}, 2047 (1993).

\bibitem{Maartens} R. Maartens, Class. Quant. Grav. \textbf{12}, 1455 (1995).

\bibitem{Zimdahl} W. Zimdahl, Mon. Not. R. Astron. Soc. \textbf{280}, 1239 (1996).

\bibitem{Maartens1996} R. Maartens, \textit{Causal thermodynamics in relativity}, arXiv:astro-ph/9609119 (third chapter).

\bibitem{Gonzalez} N. Cruz, E. Gonz\'alez and G. Palma, Gen. Rel. Grav. \textbf{52}, 62 (2020).

\bibitem{Weinberg1971} S. Weinberg, Atrohpys. J. {\bf{168}}, 175 (1971).

\bibitem{Cornejo} O. Cornejo-P\'erez and J. A. Belinch\'on, Int. J. Mod. Phys. D \textbf{22}, 1350031 (2013).

\bibitem{Harko} M. K. Mak and T. Harko, Int. J. Mod. Phys. D \textbf{13}, 273 (2004).

\bibitem{Harko1998} M. K. Mak and T. Harko, Gen. Rel. Grav. \textbf{8}, 1171 (1998).

\bibitem{Chimento1998} L. P. Chimento, A. S. Jakubi, V. M\'endez, Int. J. Mod. Phys. D \textbf{7}, 177 (1998).

\bibitem{Chimento} L. P. Chimento, A. S. Jakubi and D. Pav\'on, Int. J. Mod. Phys. D \textbf{9}, 43 (2000).

\bibitem{Brevik} I. Brevik, O. Gron, J. de Haro, S. D. Odintsov and E. N. Saridakis, Int. J. Mod. Phys. D 26, 1730024 (2017).

\bibitem{Maartens1997} R. Maartens and V. M\'endez, Phys. Rev. D \textbf{55}, 1937 (1997).

\bibitem{Chimento1} L. P. Chimento, A. S. Jakubi and V. M\'endez, Class. Quant. Grav. \textbf{14}, 3363 (1997).

\bibitem{Cruzphantom} M. Cruz , N. Cruz and S. Lepe, Phys. Lett. B \textbf{769}, 159 (2017).

\bibitem{Beesham} G. Acquaviva and A. Beesham, Phys. Rev. D \textbf{90}, 023503 (2014).

\bibitem{Beesham1} G. Acquaviva and A. Beesham, Class. Quant. Grav. \textbf{32}, 215026 (2015)

\bibitem{Cruz2018} N. Cruz, E. Gonz\'alez, S. Lepe and D. S\'aez-Chill\'on G\'omez, J. Cosmol. Astropart. Phys. \textbf{12}, 017 (2018).

\bibitem{Hiscock} W. A. Hiscock and L. Lindblom, Ann. Phys. (N.Y.) \textbf{151}, 466 (1983).

\bibitem{Mathew2017} J. Mohan N. D., A. Sasidharan and T. K. Mathew, Eur. Phys. J. C \textbf{77}, 849 (2017).

\bibitem{Piattella} O. Piattella, J. Fabris, W. Zimdahl, J. Cosmol. Astropart. Phys. \textbf{05}, 029 (2011).

\bibitem{Cruz2017} N. Cruz and S. Lepe, Phys. Lett. B \textbf{767}, 103 (2017).

\bibitem{Cruz2017a} M. Cruz , N. Cruz and S. Lepe, Phys. Rev. D \textbf{96}, 124020 (2007).

\bibitem{Goodman} J. Goodman and J. Weare, Commun. Appl. Math. Comput. Sci. \textbf{5}, 65 (2010).

\bibitem{Foreman} D. Foreman-Mackey, D. W. Hogg, D. Lang and J. Goodman, Publ. Astron. Soc. Pac. \textbf{125}, 306 (2013).

\bibitem{Scolnic} D. M. Scolnic \textit{et al.}, Astrophys. J. \textbf{859}, 101 (2018).

\bibitem{Magana} J. Magaña, M. H. Amante, M. A. Garcia-Aspeitia and V. Motta, Mon. Not. Roy. Astron. Soc. \textbf{476}, 1036 (2018). 

\bibitem{Tripp} R. Tripp, Astron. Astrophys. \textbf{331}, 815 (1998).

\bibitem{Kessler} R. Kessler and D. Scolnic, Astrophys. J. \textbf{836}, 56 (2017).

\bibitem{Lazkoz} R. Lazkoz, S. Nesseris and L. Perivolaropoulos, JCAP \textbf{2005}, 010 (2005).

\bibitem{Jimenez} R. Jimenez and A. Loeb, Astrophys. J. \textbf{573}, 37 (2002).

\bibitem{Riess2016} A. G. Riess \textit{et al.}, Astrophys. J. \textbf{826}, 56 (2016).

\bibitem{Schwarz} G. Schwarz, Ann. Stat. \textbf{6}, 461 (1978).

\bibitem{Brevik11} B.D. Normann and I. Brevik, Entropy, \textbf{18}, 215 (2016).

\bibitem{Cruzpowerlaw} N. Cruz, A. Hernández-Almada and O. Cornejo-Pérez, Phys. Rev. D \textbf{100}, 083524 (2019).

\end{thebibliography}
\end{document}